\documentclass[11pt,a4paper]{article}
\usepackage[utf8]{inputenc}
\usepackage[english]{babel}
\usepackage{amsmath}
\usepackage{amsfonts}
\usepackage{amssymb}
\usepackage{graphicx}
\usepackage{a4wide}
\usepackage{xspace}

\graphicspath{{figures/}}

\usepackage{float}
\usepackage{tikz}
\usepackage{physics}
\usepackage{subcaption} % Used to put 2(or more) figures next to each other
\usepackage{xcolor}     % Used for colored text
\usepackage{booktabs}   % Used for tables generated by pandas
\usepackage{tabu}       % Used to define new column types
\usepackage[
    backend=biber, style=phys
]{biblatex}
\usepackage{hyperref}
\newcolumntype{L}{>{$}l<{$}}
\newcolumntype{C}{>{$}c<{$}}
\newcolumntype{R}{>{$}r<{$}}

\hypersetup{
    colorlinks,
    linkcolor={red!50!black},
    citecolor={green!50!black},
    urlcolor={blue!80!black}
}

\newcommand{\gammapoint}{$\Gamma$-point }

\newcommand{\defectcharge}[2]{\text{#1}(#2)}

\newenvironment{rcases}
  {\left.\begin{aligned}}
  {\end{aligned}\right\rbrace}

\title{The charge cycle of group IV vacancy centers in diamond:\\ From DFT to rate equations}
\author{Joshua Claes$^1$, Bart Partoens$^1$, Dirk Lamoen$^2$\\ 
\small$^1$ COMMIT, Physics Department, University of Antwerp. Groenenborgerlaan 171, 2020 Belgium \\
\small $^2$ EMAT, Physics Department, University of Antwerp. Groenenborgerlaan 171, 2020 Belgium}

\begin{document}
\maketitle

\begin{abstract}
The silicon vacancy center in diamond is a promising system for quantum technologies due to its exceptional optical and spin properties. This has led to great interest in the silicon-vacancy center as well as in the other group IV vacancy centers. In this work, we model the charge cycle of the group IV vacancy centers from the $-2$ to $0$ charge state. As a first step, we compute the onset energies for all relevant one- and two-step ionization processes. Based on these results, we then derive the rate equations using Fermi's golden rule.
\end{abstract}

\section{Introduction}
Group IV vacancy centers in diamond are point defects consisting of a group IV element (Si, Ge, Sn, or Pb) situated between two vacant carbon lattice sites, often referred to as a "split-vacancy" configuration. Due to their symmetry $D_{3d}$ symmetry with an inversion center, . Among these, the silicon vacancy (SiV) is the most extensively studied and popular center due to its promising optical properties. In its negative charge state, the SiV exhibits narrow optical emission with 70\% of the emission from in the zero phonon line \cite{Sipahigil2014, Neu_2011}. However, this state is associated with a short spin coherence time due to the dynamic Jahn-Teller effect \cite{Rogers2014,Jahnke_2015}. In contrast, in the neutral charge state, the SiV demonstrates significantly improved spin coherence, with coherence times extending up to 1 second at cryogenic temperatures \cite{Goss1996, Goss2005,rose2018observation}. The other group IV vacancies are expected and have been shown to exhibit similar properties as a result of their similarity in the atomic structure.

To get these defects in the desired charge state, for example the neutral charge state with the longer coherence time, one needs to shift the Fermi energy by doping the diamond structure. Alternatively the charge state of defect can be altered by exiting an electron from the defect to the conduction band or from the valence band to the defect. This could be done to prepare the defect in a desired state or for example as part of PDMR experiment \cite{PDMR_Bourgeois_Nesladek}. To alter the SiV charge state from negatively charge to neutral a UV excitation is needed \cite{Charge_state_dyn_Nitrogen, Gali_thiering_2018_g4v}. 

In this study, we investigate the charge transitions and the cycling between the $-2$, $-1$, and $0$ charge states of group IV vacancy centers. This is achieved by calculating the onset energies and finding rate equations to model the concentration of different charge states.

During the study of this work some similar works have been released \cite{Charge_state_dyn_Nitrogen, ChargeTrans_SnV}. Although these work are of a more experimental nature and focus one particular group IV defect.

\section{Methodology \& theory}

\subsection{Group theory description}
The group IV vacancy centers consist of a double vacancy with the IV element at the center of the two vacant site in the diamond lattice, resulting in the $D_{3d}$ symmetry group. Through a group theory analysis it can be derived that the dangling bonds of the divacancy from orbitals which are part of the $a_{1g}$, $a_{2u}$, $e_u$ and $e_g$ irreducible representation of the $D_{3d}$ \cite{Coulson_divacancy}. The group IV element has an $sp^3$, which will reduce to the $a_{1g}$, $a_{2u}$ and the $e_u$ under the $D_{3d}$ crystal field \cite{GaliMaze2013, Haenens_Johansson2011}. The $a_{1g}$, $a_{2u}$ and $e_u$ of the dangling carbon bonds, the group IV element may combine but the $e_g$ should be pure carbon. The bonding, anti-bonding  combinations of these states form the defect levels of the group IV vacancy centers \cite{GaliMaze2013}.  

In the neutral charge state, the group IV vacancy center contains 10 electrons, with 6 originating from the dangling carbon bonds and 4 from the group IV element. This leads to a single-electron configuration of $a_{1g}^2 a_{2u}^2 e_u^4 e_g^2$. We determine the multi-electron ground state, by investigate the complementary hole system $e_g^2=e_g \otimes e_g$. By projecting this product state on the irreducible representations of the $D_{3d}$ group we find 3 possible multi-electron states, one of which is a triplet and is depicted in equation \eqref{eq:SiV0_gs}.
\begin{equation}
    \ket{^3A_{2g}} =  \mathcal{A} \ket{e^x_g e^y_g}\otimes \begin{cases}
 \ket{\uparrow \uparrow} \\ 
 \mathcal{S}\ket{\uparrow \downarrow} \\ 
 \ket{\downarrow \downarrow} 
 \end{cases} \label{eq:SiV0_gs}
\end{equation}
Here we have introduced the symmetry operator $\mathcal{S}\ket{ab} = \frac{1}{\sqrt{2}} (\ket{ab}+\ket{ba})$ and the asymmetry operator $\mathcal{A}\ket{ab} = \frac{1}{\sqrt{2}} (\ket{ab} - \ket{ba})$, following the notation in \cite{Thiering2019}.
We note that there are also multiple singlet with this single-electron configuration\cite{Thiering2019}. However these singlets are higher in energy and not relevant for the discussion a head.  The excite state of the neutral group IV vacancies is quite complicated. In the single electron picture and electron goes from  $e_u$ to an $e_g$ orbital, leaving the defect in the $a_{1g}^2 a_{2u}^2 e_u^3 e_g^3$ state. By calculating $e_u \otimes e_g$ the multi particle states we're determined to be
\begin{equation}
    \begin{rcases}
        \ket{^3A_{2u}} &= \frac{1}{\sqrt{2}} \mathcal{A}(\ket{e^x_u e^x_g } + \ket{e^y_u e^y_g})\\
        \ket{^3E_{ux}} &= \frac{1}{\sqrt{2}} \mathcal{A}(\ket{e^x_u e^x_g } - \ket{e^y_u e^y_g})\\
        \ket{^3E_{uy}} &= \frac{1}{\sqrt{2}} \mathcal{A}(\ket{e^x_u e^y_g } + \ket{e^y_u e^x_g})\\
        \ket{^3A_{1u}} &= \frac{1}{\sqrt{2}} \mathcal{A}(\ket{e^x_u e^y_g } - \ket{e^y_u e^x_g})
    \end{rcases} 
    \otimes \begin{cases}
 \ket{\uparrow \uparrow} \\ 
 \mathcal{S}\ket{\uparrow \downarrow}\\ 
 \ket{\downarrow \downarrow} 
 \end{cases} 
\end{equation}
However, in \cite{Thiering2019} it is shown that actual excited state is a product Jahn-Teller state with the following configuration.
\begin{equation}
    \ket{^3\Tilde{A}_{2u}} =\frac{1}{\sqrt{2}} \ket{^3A_{2u}} - \frac{\cos(\phi)}{\sqrt{2}}\ket{^3E_{ux}} - \frac{\sin(\phi)}{\sqrt{2}}\ket{^3E_{uy}}
\end{equation}

In the negative charge state an additional electron is added to our single-electron configuration and thus we find $a_{1g}^2 a_{2u}^2 e_u^4 e_g^3$. For this single-electron configuration it simple to find the multi-electron state by looking at the complementary hole system, which consist of a single orbital $e_g$. Therefor, the multi-electron system will be $\ket{^2E_g}$. To excite the defect in this state, an electron from a $e_u$ orbital must be promoted to the $e_g$ orbital and thus the defect goes from $\ket{^2E_g}$ to $\ket{^2E_u}$.

By filling the last remaining orbital, the double negatively charged group IV vacancy center is obtained, with the $a_{1g}^2 a_{2u}^2 e_u^4 e_g^4$  single electron configuration. 
% In this state the defect has the $\ket{^1A_{2u}}$ multi-electron states. 
For the double negative group IV vacancy there is obviously now excited state, all higher unoccupied bands are in the conduction band.

\subsection{Einstein coefficients}\label{sec:Einstein_coefficients}
Although color centers are not atoms, they have an atom like character.  We therefor give a brief overview of atomic optical transitions and borrow from this well studied field.

Suppose we have an atom, which we irradiate with a monochromatic beam of N photons of energy $\hbar \omega$ per unit volume. The probability per unit time or transition rate for which the atom goes from state $i$ to state $f$  is given by Fermi's Golden rule \cite{Sakurai_modernQM, fowler_color_centers}.
\begin{equation}
    W_{if} = \frac{4 \pi^2 e^2 |\varepsilon_f - \varepsilon_i|}{3 \hbar} N |\vec{r}_{if} |^2 \delta(|\varepsilon_f - \varepsilon_i| - \hbar \omega) \label{eq:transition_rate_atom}
\end{equation}
Here $\vec{r}_{if}$ is the transition dipole moment matrix element, $\varepsilon_{i/f}$ the energy of states $i$ and $f$\footnote{We use the suggestive notation $\varepsilon$ for the total energies because later on we will look at the density of states $\rho(\varepsilon$) where $\varepsilon$ will be an eigenvalue of our DFT calculations.} and $e$ the elementary electron charge. We note that this equation is only valid for small phonon densities $N$ and both $i$ and $f$ are discrete states \cite{fowler_color_centers}.  The transition rate can be related to another quantity namely the absorption coefficient $\mu$ which defined by Eq. \eqref{eq:absorption_coeff_def}
\begin{equation}
    dI = -\mu I dx \rightarrow I = I(0) e^{-\mu x} \label{eq:absorption_coeff_def}
\end{equation}
which describe the intensity decrease $dI$ as a function of the thickness of our sample $dx$ \cite{fowler_color_centers, stoneham}.  Therefor, the absorption coefficient must be the energy removed per  unit time per unit volume of a unit intensity\cite{fowler_color_centers}.  The absorption coefficient $\mu_{if}$ of the process $i \rightarrow f$ is therefor proportional to the transition rate, however there might be multiple atoms in our unit volume there for we need to multiply $W_{ij}$ by the atom density per unit volume $N_a$ and divide by the photon flux $N c$. We then find 
\begin{equation}
    \mu_{if} = \int \frac{W_{if} N_a}{Nc} d(\hbar \omega) \label{eq:absorption_coeff} 
\end{equation}
To keep the formulation of Eq. \eqref{eq:absorption_coeff} general we add an integral of the energy spectrum of the incident light, such that all transition are included in the absorption coefficient. 
Another quantity related to $W_{if}$ and $\mu_{if}$ is the absorption cross section, which can be seen as the absorption coefficient of a single defect. 
\begin{equation}    
    \sigma_{if} = \frac{\mu_{ij}}{N_a} = \frac{4 \pi^2 e^2 \hbar \omega_{if}}{3 \hbar c} |\vec{r}_{if} |^2 = \frac{4 \pi^2 \alpha }{3} \hbar \omega_{if} |\vec{r}_{if} |^2  \label{eq:abs_cross_section}
\end{equation}
Where we used $\alpha = \frac{e^2}{\hbar c}$. Note that $\sigma_{if}$ has the dimensions energy times area. 

With the goals of deriving rate equations, we now want to find a way to express the Einstein coefficients in term of these quantities. Suppose that we had  $N_f$ atoms in the $f$ state and $N_i$ atoms in the $i$ state. The rate of change of state $f$ should be proportional to the number of particles in state $i$, $N_i$  and the energy density of electron magnetic field $\rho(\omega_0)$. The energy density $\rho(\hbar \omega)$, should be proportional $N$, to the number of photons of energy $\hbar \omega$  per unit volume as is present in the equation of the transition rate \eqref{eq:transition_rate_atom}. Similarly stimulated emission will be proportional to $N_f$ and $\rho(\omega_0)$. The spontaneous emission should be independent of the electromagnetic field and is therefor only proportional to the number atoms in state $f$.
\begin{equation}
\begin{cases}
        \frac{d N_i}{dt} & =  -\tilde{B}_{if} \rho_\gamma(\hbar \omega) N_i + \tilde{B}_{fi}  \rho_\gamma(\hbar \omega) N_f + A_{fi} N_f\\
        \frac{d N_f}{dt} &= +\tilde{B}_{if} \rho_\gamma(\hbar \omega) N_i - \tilde{B}_{fi}  \rho_\gamma(\hbar \omega) N_f - A_{fi} N_f \\
\end{cases}
\end{equation}
With $\rho_\gamma(\hbar \omega)$ the energy density of the electromagnetic field, which should be equal to $\hbar \omega N$ if $\hbar \omega$ is the energy of the incident photons. If we assume that $\hbar\omega=\hbar \omega_{if} = \varepsilon_f - \varepsilon_i$ then the coefficients $\tilde{B}_{if} \rho_\gamma(\hbar \omega_{if})$ should be the transition probability of state $i$ to all states $f$ i.e. $W_{i,f}$ and have unit $1/s$. Thus $\tilde{B}_{if} \rho_\gamma(\hbar \omega_{if})$ is exactly the earlier described transition probability $W_{if}$ after we take in to account all similar final states $[f]$ by integrating over $\int \rho(\varepsilon_f) d\varepsilon_f$ with $\rho(\varepsilon)$ the electronic density of states. Note that the unit of the density of states should be states per energy. We now define the actual Einstein coefficients as
\begin{equation}
\begin{cases}
        \frac{d N_i}{dt} &= -B_{if} \rho(\varepsilon_{f}) \rho_\gamma(\hbar \omega_{if}) N_i + B_{fi} \rho(\varepsilon_{i}) \rho_\gamma(\hbar \omega_{if}) N_f + A_{fi} N_f \\
        \frac{d N_f}{dt} &= +B_{if} \rho(\varepsilon_{f}) \rho_\gamma(\hbar \omega_{if}) N_i - B_{fi} \rho(\varepsilon_{i}) \rho_\gamma(\hbar \omega_{if}) N_f - A_{fi} N_f
\end{cases} \label{eq:rate_eq_general}
\end{equation}
with 
\begin{equation}
    B_{if} = \frac{4 \pi^2}{3} \alpha c |\vec{r}_{if} |^2.
\end{equation}
The coefficient for the spontaneous emission is given by \cite{MarkFox_optical_properties} 
\begin{equation}
    A_{fi} = \frac{2 \hbar \omega^3}{\pi c^3} B_{fi}
\end{equation}
 
\subsubsection{The effect of the medium}
Since we are studying defects within a material, we must account for the influence of the surrounding medium on our system. This includes modifying the speed of light as $c\rightarrow c/n$ where $n$ is the refractive index of the material, and recognizing that the effective electric field at the defect site  $E_{eff}$, may differ from the externally applied field, $E_0$ \cite{fowler_color_centers, stoneham}. For both the optical cross section and the Einstein coefficient $B$, we must apply a correction factor:
\begin{equation}
    \frac{1}{n} \left(\frac{E_{eff}}{E_0}\right)^2
\end{equation}
while the spontaneous emission needs to be multiplied by
\begin{equation}
    n \left(\frac{E_{eff}}{E_0}\right)^2.
\end{equation}

There are many ways in which the surrounding medium affects the defect center, and these influences significantly impact the center's overall properties. However, they do not directly alter the form of the transition rate or optical cross section. Instead, these effects shape the broader analysis of the center. Below, we outline several important ways in which the medium influences defect behavior.

One of the primary effects is the electrostatic potential that the medium exerts on the center. This affects both the structural properties and the charge state of the defect. Structural effects are typically already captured in DFT calculations, which are our main tool for extracting defect parameters. The defect’s charge state is usually chosen, at least initially, when setting up the rate equations, but it can also be analyzed based on formation energies obtained from DFT.

Related to the potential effect, the crystal also imposes a crystal field with a specific symmetry on the defect center. This generally lifts certain degeneracies in the electronic structure. Additionally, the incorporation of a defect may further lower the symmetry of the host crystal. These symmetry-related effects are also included in DFT calculations. Furthermore, group theory analysis based on the symmetry of the defect can be used to derive many basic properties without the need for detailed calculations \cite{fowler_color_centers, stoneham}.

Finally, the defect center interacts with both the crystal’s lattice vibrations and its own localized vibrational modes. These interactions are responsible for phonon sidebands observed in emission spectra.  The effect of the vibrational broadening for the ZPL emissions and the photo-ionization is introduced in our calculation by following \cite{Razinkovas_photoionization, Alkauskas_2014, Razinkovas_Vibrational_NV}.

\subsection{The rate equations for the group IV charge cycle}
At first glance, these coefficients suggest that we can directly simulate the time evolution of the neutral and negative charge state populations. However, the system is more complex than it appears.  Specifically, the Einstein coefficients describe optical transitions rather than charge state changes. 
We now use the example of the rate equations for the system in the case where only the negative an neutral charge state exist and there are only transitions between the ground state to explain our model. Here $n_0$ and $n_{-1}$ represent the concentration of the neutral and negative charge state respectively.
If a defect initially in the negative charge state absorbs a photon and transitions to the neutral state, one might expect it to return to the negative state via spontaneous or stimulated emission. However, it is clear that once the laser is turned off, spontaneous emission alone cannot restore the defect to its original charge state. Moreover, under the current assumptions, the rate of stimulated emission should be equal to the rate of absorption from the neutral to the negative state. This is unlikely, as the two processes have different onset energies. This discrepancy suggests that both stimulated and spontaneous emission should, in principle, be absent in the system as described. However, it is reasonable to expect that for a brief period following absorption, both emission processes could still occur. 

To address this, we introduce two new auxiliary states: $n^{e^-}_0$ and $n^{h^+}_{-1}$. Here, $n^{e^-}_{0}$ represents the defect in the neutral state accompanied by an electron in the conduction band ($\varphi_c$), while $n^{h^+}_{-1}$ represents the defect in the negative state accompanied by a hole in the valence band ($\varphi_v^h$). From these intermediate states, both stimulated and spontaneous emission processes are possible. To connect these new states to the original ground states, we introduce a transition parameter $C_{i,j}$, which describes how quickly the system relaxes from the auxiliary "excitonic" states to their respective ground states, where the electron or hole has moved away from the defect. By choosing $C_{i,j}$ sufficiently large, we recover a regime where emission is effectively suppressed, consistent with our initial observation. The system of equations governing the transitions between the neutral and negative ground states is given in Eq. \eqref{eq:system_of_eq_n0_n-1}.

In the absence of all transition expect the decay of the "excitonic" state, the coefficient $C$ is the inverse of the lifetime $\tau$ of this state \cite{MarkFox_optical_properties}. 
\begin{align}
    \frac{dn^{h^+}_{-1}}{dt}    &= - C_{0,-1} n^{h^+}_{-1} \\
    \Rightarrow n^{h^+}_{-1}    &= n^{h^+}_{-1}(0) e^{-C_{0,-1} t} \\
                                &= n^{h^+}_{-1}(0) e^{-t/\tau}
\end{align}
This give us the following set of equations.
\begin{equation}
    \left\{ 
    \begin{array}{rrrrr} 
         \frac{d n_0}{dt}          =& - B_{0,-1}  \rho_0(0) n_0    &+ B_{0, -1}  \rho_0(-\varepsilon_\gamma)n^{h^+}_{-1} &+ A_{-1,0} n^{h^+}_{-1} &+ C_{-1, 0} n^{e^-}_0\\
         \frac{dn^{h^+}_{-1}}{dt}  =&   B_{0,-1}  \rho_0(0) n_0    &- B_{0, -1}  \rho_0(-\varepsilon_\gamma) n^{h^+}_{-1} &- A_{-1,0} n^{h^+}_{-1} &- C_{0,-1} n^{h^+}_{-1} \\
         \frac{d n_{-1}}{dt}       =& - B_{-1,0} \rho_{-1}(\varepsilon_{\gamma}) n_{-1} &+ B_{-1,0}  \rho_{-1}(0) n^{e^-}_{0}   &+ A_{0,-1} n^{e^-}_{0}  &+   C_{0,-1} n^{h^+}_{-1}\\
         \frac{d n^{e^-}_{0} }{dt} =&   B_{0,-1} \rho_{-1}(\varepsilon_{\gamma})n_{-1}  &- B_{-1,0}  \rho_{-1}(0) n^{e^-}_{0}   &- A_{0,-1} n^{e^-}_{0}  &-  C_{-1, 0} n^{e^-}_{0} 
    \end{array} \right.\label{eq:system_of_eq_n0_n-1}
\end{equation}
Here we define $ \rho_{q}(\varepsilon)$ as the density of states for the system in charge state $q$ at energy  $\varepsilon$. Where the density of states is shifted such that the initial defect state is found at $\varepsilon_i=0$ and the density of the final state is found at $\rho(\varepsilon_f)$ with $\varepsilon_f$ equal to the energy of the incident light\footnote{Usually this will be calculated using DFT and therefor the calculated onset energy will not necessarily be the same as the gap between a defect level and the valence/conduction band. There a scissor operator is applied such that the density of state respects the onset energy. In other words if $r_{if}(\varepsilon)$ is on all parts of the bands which fall in an energy interval $\delta \varepsilon$ the the density of states will correspond to that of the energy interval $\delta\varepsilon$}. In the case of excitation where a hole is create, like the excitation $n_0$ to $n_{-1}^{h+}$ in Eq. \eqref{eq:system_of_eq_n0_n-1}, from the perspective of the excited electron the defect levels are the final states and a valence band is the initial state. Therefor, the density $\rho_0(0)$ appears as the density of the final state in Eq. \eqref{eq:system_of_eq_n0_n-1} because it is the final state. For the initial state, or the final states in the stimulated emission process, we added a minus sign to indicate that the density is obtained from the valence bands. 

The equations as described above only focus on the single electron picture. However the defect states are multi electron states and as determined in the appendix this sometimes results in an extra "degeneracy factor" due to multiple single electron states that can undergo the transition, as is the case for the $Q=-2$ state or an initial state can transition to different orbital or spin states of the multi-particle final states. Therefor, some transitions have an additional degeneracy factor.

For each transition in Figure \ref{fig:charge_cycle_scheme_SiV} that involves a change in charge state, we can formulate equations similar to those in \eqref{eq:system_of_eq_n0_n-1}. 

\section{The degenarcy factor and onset energies of the photo-ionization}
\subsection{Ground state photo-ionization}
The simplest optical process for changing the charge state is one-step photoionization, where an electron is either directly added to or removed from the ground state. The onset energy for this process is referred to as the adiabatic or optical charge transition energy. This transition energy is determined by finding the intersection of the formation energies $E^f(q)$ and $E^f(q \pm 1)$ \cite{First_prin_def_solid}. The formation energy $E^f$ of a defect $D$ in charge state $q$ can be calculated using Eq. \eqref{eq:Eform}.

\begin{equation}
E^f(D^q) = E_{tot}(D^q) - E_{tot}(\text{bulk}) - \sum_i \mu_i n_i + q(E_{VBM}+\varepsilon_F+\Delta V) + E_{corr}\label{eq:Eform}
\end{equation}

Here, $E_{tot}$ represents the total energy, $\mu_i$ is the chemical potential of a removed or added atom $i$, and $n_i$ is the number of atoms removed or added\footnote{Given that we're only interested in the transition energies the term with $n_i$ and $\mu_i$ does not need to be added to the formation energy}. $E_{VBM}$ is the energy of the valence band maximum (VBM) in the calculation, $\varepsilon_F$ is the Fermi energy, and $\Delta V$ accounts for the shift in the electrostatic energy potential. The value of $\Delta V$ is determined by uniformly sampling the electrostatic potential across the defect and pristine supercells. For each sample point, the difference in electrostatic potential is calculated and placed into a histogram. $\Delta V$ is then set to the value corresponding to the largest bin in the histogram. Lastly, $E_{corr}$ is a correction term accounting for interactions between the defect and its periodic mirror images as well as the uniform charged background, we use the Makov Payne correction \cite{Makov_Payne_corr}.

If the transition energy is calculated as described above, the result is the thermal charge transition energy, as it accounts for the relaxation of the final state's structure. However, during an optical transition, the structure does not have sufficient time to relax. As a result, after the transition, the structure remains that of the initial charge state. To determine the correct optical transition energy, the formation energy of the final state is calculated using the geometry of the initial state \cite{Walle2004FirstprinciplesCF, First_prin_def_solid}.

We will now investigate the optical cross section of one-step photo-ionization. Because DFT results are in the single-electron picture, we need to find a connection between the transition dipole moment of the single-particle state and that of the multi-electron state. 

\subsubsection{Photo-ionization of the $Q=-2$ state}
We start by investigate the lowest charge state, the double negative group IV vacancy center. In this charge state the defect has the single-electron state $ \ket{e_g^x \bar{e}_g^x e_g^y \bar{e}_g^y}$, where only explicitly write the 4 highest energy $e_g$ states to keep the notation compact and use $\bar{e}$ to denote the spin down orbitals.. Due to symmetry the four $e_g$ levels have the same energy, we now illustrate how to calculate the transition dipole moment for the case where the $\bar{e}_g^y$ is excited to he conduction band $\bar{\varphi}_c$, following \cite{Razinkovas_photoionization}. Here we assume that all single-electron orbitals are the same in the initial and final state, therefor the initial and final state only differ by the occupied $\bar{e}_g^y$ and $\bar{\varphi}_c$ level.
\begin{align}
    \vec{r}_{ij}    &= \bra{e_g^x \bar{e}_g^x e_g^y \bar{e}_g^y} \hat{O} \ket{^2E_g \otimes \bar{\varphi}_c} \\
                    &= \bra{e_g^x \bar{e}_g^x e_g^y \bar{e}_g^y} \hat{O} \ket{e_g^x \bar{e}_g^x e_g^y\otimes \bar{\varphi}_c} \\
                    &= \bra{\bar{e}_g^y} \hat{r} \ket{\bar{\varphi}_c} \label{eq:rij_Q-2_to_Q_-1}
\end{align}
Eq. \eqref{eq:rij_Q-2_to_Q_-1} gives the matrix element for one of four possible transitions. Therefor, the optical cross section of this transition needs to be multiplied by a factor 4.

\subsubsection{Photo-ionization of the $Q=-1$ state}
In the negative charge state the group IV vacancy center is in the $\ket{^2E_g}$, suppose the defect starts in the $m_s=1/2$ state. Then the single-electron configuration could be written as $\ket{e_g^x e_g^y \bar{e}_g^{\, x/y}}$. 
From this state there are 2 possible photo-ionization channels, one to the $Q=-2$ charge state and one to the $Q=0$ neutral state. 

To excite this state to the $Q=-2$ state an electron needs to be excited from the valence bands to the empty $\bar{e}_g^{\,y/x}$ defect level. In this case there is only one final state the $Q=-2$ ground state $\ket{e_g^x \bar{e}_g^x e_g^y \bar{e}_g^y}$ with a hole $\bar{\varphi}_v^h$ in the valence band. By assuming again that that the many electron wavefunction of the initial and final state differ only in the occupation of the $\bar{e}_g^{\,y/x}$ level and the hole $\bar{\varphi}_v^h$. We find that
\begin{align}
    \vec{r}_{ij}    &=\bra{^2E_g^y,1/2} \hat{O} \ket{^1A_{1g}\otimes \bar{\varphi}^h_v}\\
                    &=\bra{e_g^x \bar{e}_g^{x} e_g^y} \hat{O} \ket{e_g^x \bar{e}_g^x e_g^y \bar{e}_g^y \otimes \bar{\varphi}_v^h} \\
                    &= \bra{\bar{\phi}_v} \hat{r} \ket{\bar{e}_g^y}
\end{align}
Where we chose $\ket{^2E_g^y,1/2}$ as the initial state. A similar result can be found for the other three initial states. 

From the transition to the neutral charge state the process is a bit more complex. We start in the same initial $m_s=1/2$ spin state. However the final $\ket{^3A_{2g}}$ state, is triplet and thus there three different final states. Although one of these final states, the $m_s=-1$ can be excluded because this state can't be reached by exciting a single electron. From our DFT calculation we derive that the highest energy occupied $e_u$ orbital is the spin-down or minority spin, $\bar{e}_u^{\, x/y}$ orbital. If this orbital is excited the final state will be the $\ket{^3A_{2g}, 1}$. An thus this process is related to the calculate onset energy. We find
\begin{equation}
    \vec{r}_{ij} = \bra{^2E_u,1/2} \hat{O} \ket{ (^3A_{2g},1) \otimes \varphi_c}  = \bra{\bar{e}_u^{\, x/y}} \vec{r} \ket{\bar{\varphi}_c}
\end{equation}
For the transition to the $\ket{^3A_{2g}, m_s = 0}$ one of the $e_u^{x/y}$ orbitals needs to be excited to the conduction band. We assume that our initial state is $\ket{^2E_u^x, 1/2}=\ket{e_g^x \bar{e}_g^y e_g^y}$. The transition dipole moment then becomes
\begin{align}
    \vec{r}_{ij}    &= \bra{^2E_g,1/2} \hat{O} \ket{^3A_{2g},0}\\
                    &= \bra{e_g^x \bar{e}_g^y e_g^y} \hat{O} \left[ \ket{e_g^x e_g^y \otimes \varphi_c} \otimes \mathcal{S} \ket{\uparrow \downarrow} \right]\\
                    &= \bra{e_g^x \bar{e}_g^y e_g^y} \hat{O} \frac{1}{\sqrt{2}} \left[  \ket{e_g^x \bar{e}_g^y\otimes \varphi_c} +  \ket{\bar{e}_g^x e_g^y\otimes \varphi_c} \right] \label{eq:rij_Q-1_2Eg_to_Q0_3A2g_ms0_step1}
\end{align}
We now use assume that all single-electron orbitals are the same in the initial and final state, therefor the initial and final state only differ by the occupied $e_g^y$ and $\varphi_c$ level. This allows us to simplify \eqref{eq:rij_Q-1_2Eg_to_Q0_3A2g_ms0_step1} to
\begin{align}
    \vec{r}_{ij}    =& \frac{1}{\sqrt{2}} \left( \braket{e_g^x \bar{e}_g^y}{e_g^x \bar{e}_g^y} \bra{e_g^y} \hat{r} \ket{\varphi_c} +   \bra{e_g^x \bar{e}_g^ye_g^y} \hat{O} \ket{\bar{e}_g^x e_g^y \varphi_c} \right. \label{eq:rij_Eg_to_3A2g_ms0}\\
                    % =& \frac{1}{\sqrt{2}} \left( \braket{e_g^x \bar{e}_g^y}{e_g^x \bar{e}_g^y} \bra{e_g^y} \hat{r} \ket{\varphi_c} +   \braket{e_g^x \bar{e}_g^y}{\bar{e}_g^x e_g^y} \bra{e_g^y} \hat{r} \ket{\varphi_c} \right. \\
                    =& \frac{1}{\sqrt{2}} \bra{e_g^y} \hat{r} \ket{\varphi_c} \label{eq:rij_Q-1_2Eg_to_Q0_3A2g_ms0}
\end{align}
Where the last state in Eq. \eqref{eq:rij_Eg_to_3A2g_ms0} is equal to zero 
For each of the 4 initial states, taking in to account $m_s=-1/2$ to $m_s=0$ transition, we find a formula for the a matrix element $\vec{r}_{ij}$ similar to \eqref{eq:rij_Q-1_2Eg_to_Q0_3A2g_ms0}. 

In this derivations of the transitions to the $Q=-2$ and $Q=0$  we neglected the Jahn-Teller effect, our calculation indicate that the degeneracy of the $e_u^{x/y}$ was lifted due to the Jahn-Teller effect caused by the half occupied $\bar{e}_u^{\, x/y}$. This means one of the 4 initial states has lower energy, this is the state we find in our DFT calculation and for which we determine the matrix element. 

\subsubsection{Photo-ionization of the Q=0 state}
Lastly we calculate the transition dipole moment between the $Q=0$ and the $Q=-1$ state. We start by determining the transition from the $\ket{^3A_{2g}, 1}$ to the $\ket{ (^2E_g^y,1/2)}$ state.
\begin{align}
    \vec{r}_{ij}    &= \bra{^3A_{2g}, 1} \hat{O} \ket{ (^2E_g^y,1/2) \otimes \bar{\varphi}_v^h}\\
                    &= \bra{e_g^x e_g^y} \hat{O} \ket{e_g^x \bar{e}_g^x e_g^y\otimes \bar{\varphi}_v^h} \\
                    &= \bra{\hat{\varphi}_v} \hat{r} \ket{\bar{e}_g^x } \label{eq:rij_Q0_ms1_to_Q-1}
\end{align}
Because the final state is an orbital doublet we can multiple result \eqref{eq:rij_Q0_ms1_to_Q-1} by a degeneracy factor of $g=2$ while calculating the optical cross section, to take in to account the transition to the $\ket{^2E_g^x,1/2}$ state. For the transitions with the initial state $m_s=-1$ we find the same matrix element.
In the $\ket{^3A_{2g}, 0}$ initial state consist of 2 slater determinants resulting in an extra factor $1/\sqrt{2}$ in the final results, as can be seen in Eq. \eqref{eq:rij_Q0_ms0_to_Q-1}. 
\begin{align}
    \vec{r}_{ij}    &= \bra{^3A_{2g}, 0} \hat{O} \ket{ (^2E_g^y, -1/2) \otimes \bar{\varphi}_v^h} \label{eq:rij_Q0_ms0_to_Q-1_step1}\\
                    &= \frac{1}{\sqrt{2}}\left(\bra{e_g^x \bar{e}_g^y} +  \bra{\bar{e}_g^x e_g^y} \right) \hat{O}  \ket{e_g^x \bar{e}_g^x   \bar{e}_g^y\otimes \bar{\varphi}_v^h} \\
                    &= \frac{1}{\sqrt{2}} \bra{\hat{\varphi}_v} \hat{r} \ket{\bar{e}_g^x } \label{eq:rij_Q0_ms0_to_Q-1}
\end{align}
We note that we chose the final state in Eq. \eqref{eq:rij_Q0_ms0_to_Q-1_step1} such that our results in Eq. \eqref{eq:rij_Q0_ms0_to_Q-1} is exactly that of Eq. \eqref{eq:rij_Q0_ms1_to_Q-1}. However for all final states we find a similar result. Given the possible 4 final states, we need to multiply our result by a factor 4, resulting in an overall cross section which is the same as the $m_s = \pm 1 $ states.

\subsection{Two-step photo-ionization} \label{sec:twostep_photoio}
In this section we investigate the two-step photo-ionization, this are all the processes in which light first promotes the system to an excited state, and then an electron is further excited into the conduction band.
\subsubsection{The excited Q=0 state} 
\begin{figure}[H]
\centering
\includegraphics[width=0.5\textwidth]{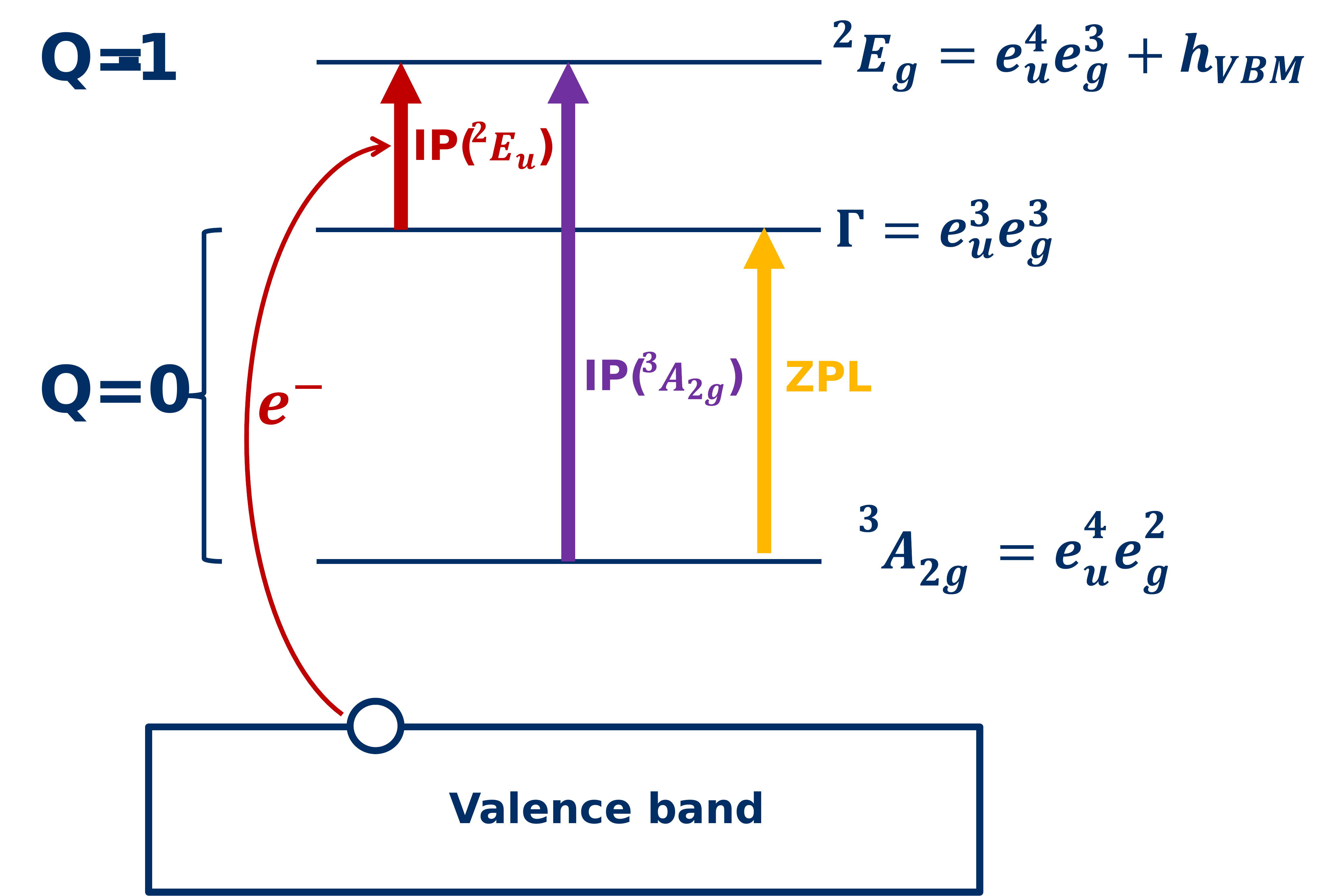}
\caption{The photoionization of the neutral state to the negatively charged state of a group IV vacancy center in diamond.}
\label{fig:scheme_photoio_0to-1}
\end{figure}

The total energy required for this process is typically lower since higher-energy levels are being filled. However, this process is restricted to the  $Q = 0$ and $Q = -1$ charge states, as the $Q = -2$ charge state has no empty levels within the band gap.
In the excited $Q = 0$ state, the defect is described by the $e^3_u e^3_g$ single-particle configuration, which corresponds to three multi-particle triplet states. Among these, the lowest-energy state is $^3A_{2g}$. Transitioning to the negatively charged state requires an electron to be excited from the valence band to the empty level within the band gap, specifically the $e_u$ level. After this transition, the defect adopts the $e^4_u e^3_g$ single-particle configuration, leaving a hole ($h_{VBM}$) in the valence band. The lowest-energy multi-particle state corresponding to the $e^4_u e^3_g$ configuration is the $^2E_g$ ground state of the negatively charged defect. The two-step ionization process is illustrated in Figure \ref{fig:scheme_photoio_0to-1}, from which Equation \eqref{eq:scheme_photoio_0to-1} can be derived to calculate the photoionization threshold for the second step. This formula neglects subtle structural differences between the ground and excited states. However, these differences are relatively minor. %To demonstrate this, we calculate the Stokes energies shown in Table \ref{tab:stokes_shift}, which are small compared to the energies required for optical transitions.
\begin{equation}
    IP(^2E_u) = IP(^3A_{2g}) - E_{ZPL} \label{eq:scheme_photoio_0to-1}
\end{equation}

With formula \eqref{eq:scheme_photoio_0to-1} the photoionization threshold is known, but to fully describe the system we also need to determine the photoionization cross section of this process. For this we need to evaluate the transition dipole moment $\vec{r_{ij}}$, which will be done using DFT. Since, DFT only has access to single particle wave function we need to find a way to connect the result of the single particle picture to realistic multi particle defect state. It has been shown by Thiering and Gali \cite{Thiering2019} that the G4V(0) excited state is a product Jahn-Teller state given by 
\begin{equation}
    \ket{^3\Tilde{A}_{2u}} = \frac{1}{\sqrt{2}} \ket{^3A_{2u}} - \frac{\cos(\varphi)}{\sqrt{2}} \ket{^3E_{ux}} - \frac{\sin(\varphi)}{\sqrt{2}} \ket{^3E_{uy}}.
\end{equation}
Using the same procedure as before we now determine the matrix element $\vec{r}_{ij}$ for the first element. Given that there are now both $e_u$ and $e_g$ are relevant, we use the more compact hole notation, denoted by the subindex $h$ for the brakets.
For now we will only focus on we will focus on evaluating $\vec{r}_{ij}$ for the first term. 
\begin{align}
    \vec{r}_{ij}    &= \bra{^3A_{2u}, m_s = 1} \hat{O} \ket{(^2E_g^y,m_s=1/2) \otimes \phi_v^h} \\
                    % &= \left(\bra{e_u^x e_g^x} + \bra{e_u^y e_g^y}  \right) \hat{O} \ket{e_u^x \bar{e}_u^x e_u^y \bar{e}_u^y e_g^x \bar{e}_g^x e_g^y \otimes \phi_v^h}
                    &= \frac{1}{\sqrt{2}} \left(\bra{\bar{e}_u^x \bar{e}_g^x}_h + \bra{\bar{e}_u^y \bar{e}_g^y}_h  \right) \hat{O} \ket{\bar{e}_g^y \otimes \phi_v^h}_h\\
                    &= \frac{1}{\sqrt{2}} \bra{\phi_v} \hat{r} \ket{e_u^y}
\end{align}
where $\phi_v^h$ represent the hole in the conduction band.

\subsubsection{From $^2E_u$ to Q=0}
We begin by describing the transition to the neutral state. The excited single-particle $Q = -1$ state is $e^3_u e^4_g$, from which one electron must be excited to the conduction band. Since the $e_g$ levels are the highest in energy, they will be the first to be excited. After this transition, the defect will be in the neutral $e^3_u e^3_g$ single-particle state, accompanied by an electron in the CBM. However, the $e^3_u e^3_g$ configuration does not correspond to the $Q = 0$ ground state; instead, it corresponds to an excited state. We assume this is the lowest-energy excited state, specifically $^3A_{2g}$. Using Figure \ref{fig:onset_-1_to_0}, we derive the following expression for the photoionization energy:

\begin{equation} 
IP(^2E_u \rightarrow Q=0) = IP(^2E_g) + E_{ZPL}^{Q=0} - E_{ZPL}^{Q=-1} 
\end{equation}

Here, $IP(^2E_g)$ is the photoionization energy from the $Q = -1$ to the $Q = 0$ ground state, and $E_{ZPL}^{Q=0/-1}$ represents the zero-phonon line of the $Q = 0$ and $Q = -1$ states.

To properly evaluate the optical cross section $\sigma$ we need to evaluate
\begin{equation}
    \bra{^2E_g} \hat{O} \ket{^3A_{2g}} = \bra{e_{g;x/y}} \hat{r} \ket{\phi_c}
\end{equation}
Our defect can either start in the $m_s=\pm\frac{1}{2}$ and end up in a $m_s\in \{-1,0,1\}$.  Suppose we start in the $m_s=\frac{1}{2}$ state. We then promote one electron from the $e_{g,\downarrow}$, we assume that $\uparrow$ is the majority spin here as this is the case in our DFT calculations, level to the conduction band.  This would leave the defect in the $m_s=1$ state of $^3A_2$.
\begin{equation}
    \vec{r}_{ij} = \frac{1}{\sqrt{2}} \bra{\bar{e}_{g;x/y}} \hat{r} \ket{\bar{\phi}_c} 
\end{equation}
To end up in the $m_s=0$ state we must promote an electron from one of 2 $e_{g,\uparrow}$ level.  To calculate this we can evaluate for $\bra{e_{g;x;\uparrow}} \hat{r} \ket{\phi_c}$. However we must also take in to account that the $\ket{^3A_2|m_s=0}$ is not represented by a single slater determinant but instead 
\begin{equation}
    \ket{^3A_2|m_s=0}=\frac{1}{\sqrt{2}} (\ket{e_{g;x}\bar{e}_{g;y}\phi_c} + \ket{\bar{e}_{g;x} e_{g;y}\phi_c})
\end{equation}
Thus 
\begin{equation}
    \vec{r}_{ij} = \frac{1}{\sqrt{2}} \bra{e_{g;x/y}} \hat{r} \ket{\phi_c}
\end{equation}
If we neglect the effect of the different spin we see that the transition to the $m_s = \pm 1$ is about a factor 2 larger than the $m_s = 0$ as this depends on $r^2_{ij}$. The procedure for the initial state with $m_s=-\frac{1}{2}$ is identical to the one described above.

\subsubsection{From $^2E_u$ to Q=-2}
\begin{figure}[H]
\centering

\begin{subfigure}{0.49\textwidth}
    \centering
    \includegraphics[width=\textwidth]{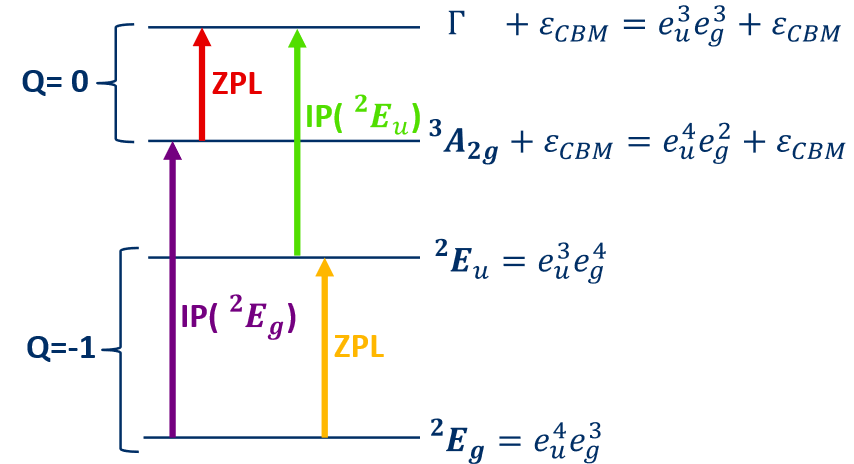}
    \caption{}
    \label{fig:onset_-1_to_0}
\end{subfigure}
\hfill
\begin{subfigure}{0.49\textwidth}
    \centering
    \includegraphics[width=0.8\textwidth]{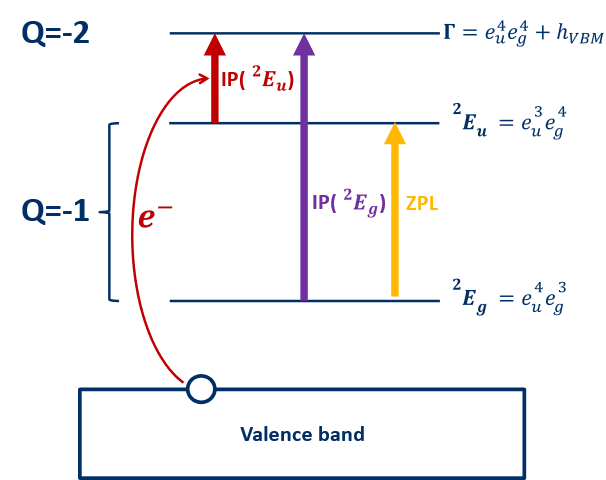}
    \caption{}
    \label{fig:onset_-1_to_-2}
\end{subfigure}
\caption{The photoionization from the negative charge state to the neutral state (a) and the double negative charge state (b) of the group IV vacancy in diamond. The state $\Gamma$ represent one of the excited states of the neutral charge state. }
\label{fig:scheme_onset_start_-1}
\end{figure}

To excite the negative $\ket{e^3_u e^4_g}$ state to the $Q = -2$ state, an electron must be excited from the valence band to the empty $e_u$ level. After this transition, all defect levels are filled, leaving only one possible state for the defect: the $\ket{e^4_u e^4_g}$, which is the $Q = -2$ ground state. Using Figure \ref{fig:onset_-1_to_-2}, we derive the photoionization threshold:
\begin{equation} 
IP(^2E_u \rightarrow Q=-2) = IP(^2E_g) - E_{ZPL}^{Q=-1} \label{eq:onset_-1_to_-2} 
\end{equation}
With Equation \eqref{eq:onset_-1_to_-2}, all photoionization thresholds between the three charge states can be determined.

\begin{figure}[H]
\centering
\includegraphics[width=\textwidth]{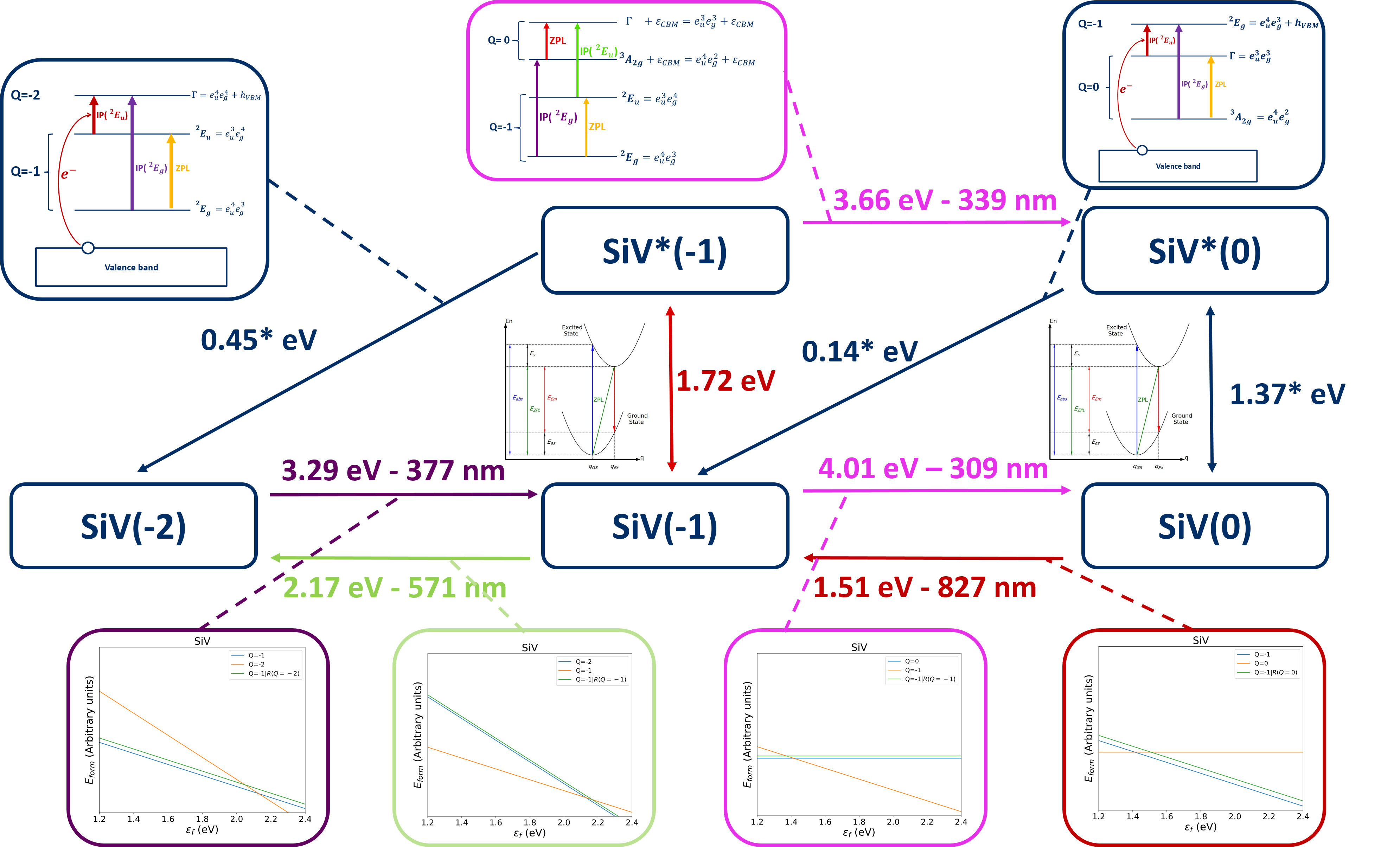}
\caption{A scheme of the charge cycle of the SiV, where each an illustration of each calculation is added.}
\label{fig:charge_cycle_scheme_SiV}
\end{figure}

\section{Computational details}

The DFT calculations were performed with the HSE06 \cite{HSE03,HSE06} hybrid functional using the (PBE) exchange correlation potential and PAW \cite{vasppaw} as implemented by the Vienna \textit{ab initio} simulation package (VASP) \cite{vaspcode1,vaspcode2,vaspcode3} taking (collinear) spin polarization in to account. All calculation where performed with an energy cutoff of $500$ eV. The lattice parameter was determined using a Birch-Murnagahan fit \cite{BMeq} with $8\times8\times8$ $k$-point grid in the Monkhorst-Pack scheme \cite{Monkhorst_Pack_scheme}. This gives us a distance between the carbon atoms and a lattice parameter of $1.537\enspace \text{\AA}$  and $3.548\enspace \text{\AA}  $, respectively,  which is in good agreement with the experimental values and similar studies \cite{semiconductor_properties, Razinkovas_photoionization}. The defect supercells were created from a $4\times4\times4$ conventional diamond supercell with 512 carbon atoms. For the relaxation of the supercell only the \gammapoint is consider for the integration over the Brillouin zone. The relaxations were stopped when all forces are below $0.01\enspace \text{eV}/\text{\AA}$. To evaluate the dipole operator $\vec{r}_{ij}$, a denser $k$-point grid is required. These calculations where done on with a $6\times6\times6$ $k$-point grid using the PBE exchange correlation functional or a correct PBE-1/2 exchange correlation functional \cite{Ferreira2008, Ferreira2011, dfthalf_NVmin}.

\section{Results}
\subsection{formation energies}
\begin{figure}[H]
    \centering
    \includegraphics[width=0.5\linewidth]{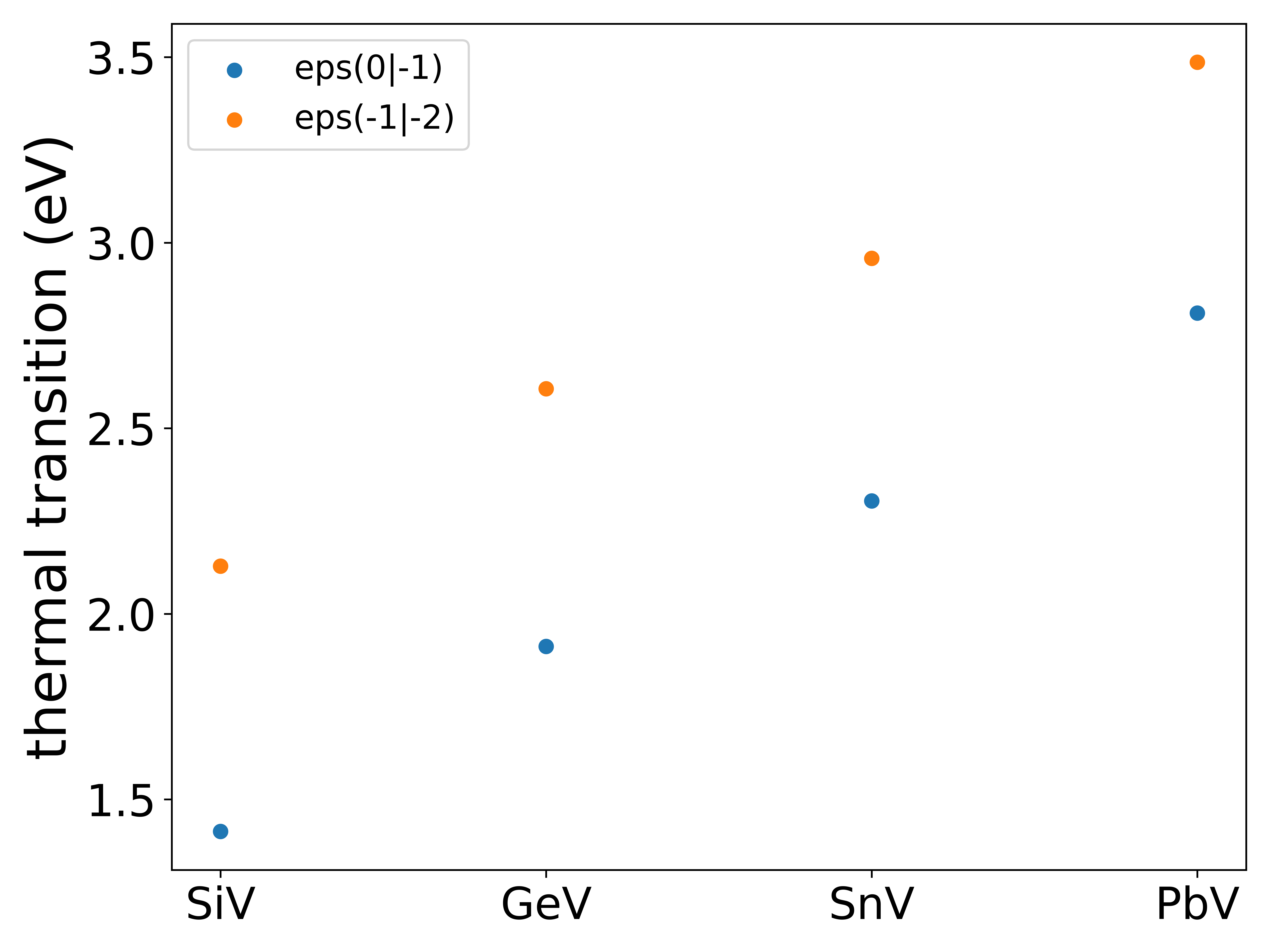}
    \caption{The transition energies for the group IV vacancy centers in diamond.}
    \label{fig:GS_photoionization}
\end{figure}

In figure \ref{fig:GS_photoionization} the photoionization threshold from the ground state is depicted for the group IV vacancy centers. The thresholds seem decrease linearly with the atomic mass of the group IV atom, these value are in line with literature \cite{Gali_thiering_2018_g4v}. 

\subsection{Group IV charge cycle diagrams}
\begin{figure}[H]
\centering

\begin{subfigure}{0.49\textwidth}
    \centering
    \includegraphics[width=\textwidth]{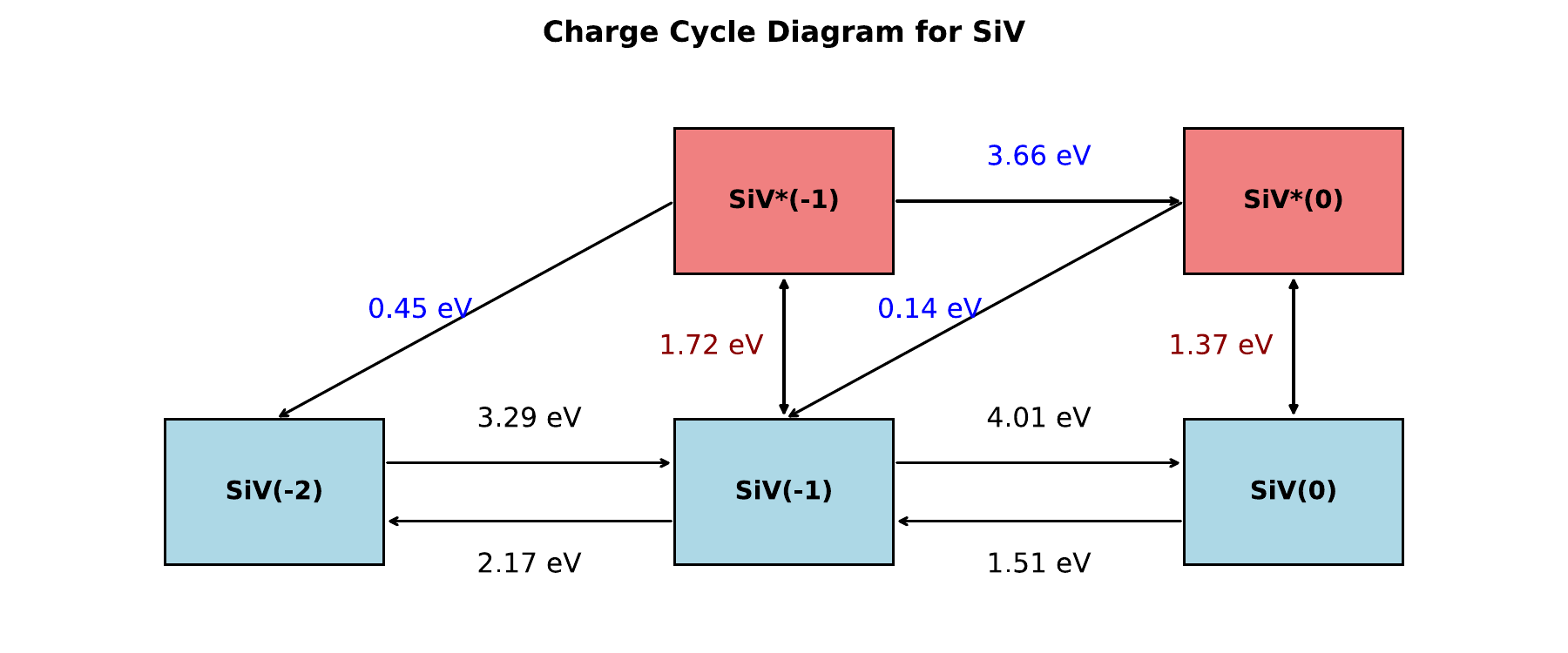}
    \caption{The onset energies for the charge cycle of SiV.}
    \label{fig:SiV}
\end{subfigure}
\hfill
\begin{subfigure}{0.49\textwidth}
    \centering
    \includegraphics[width=\textwidth]{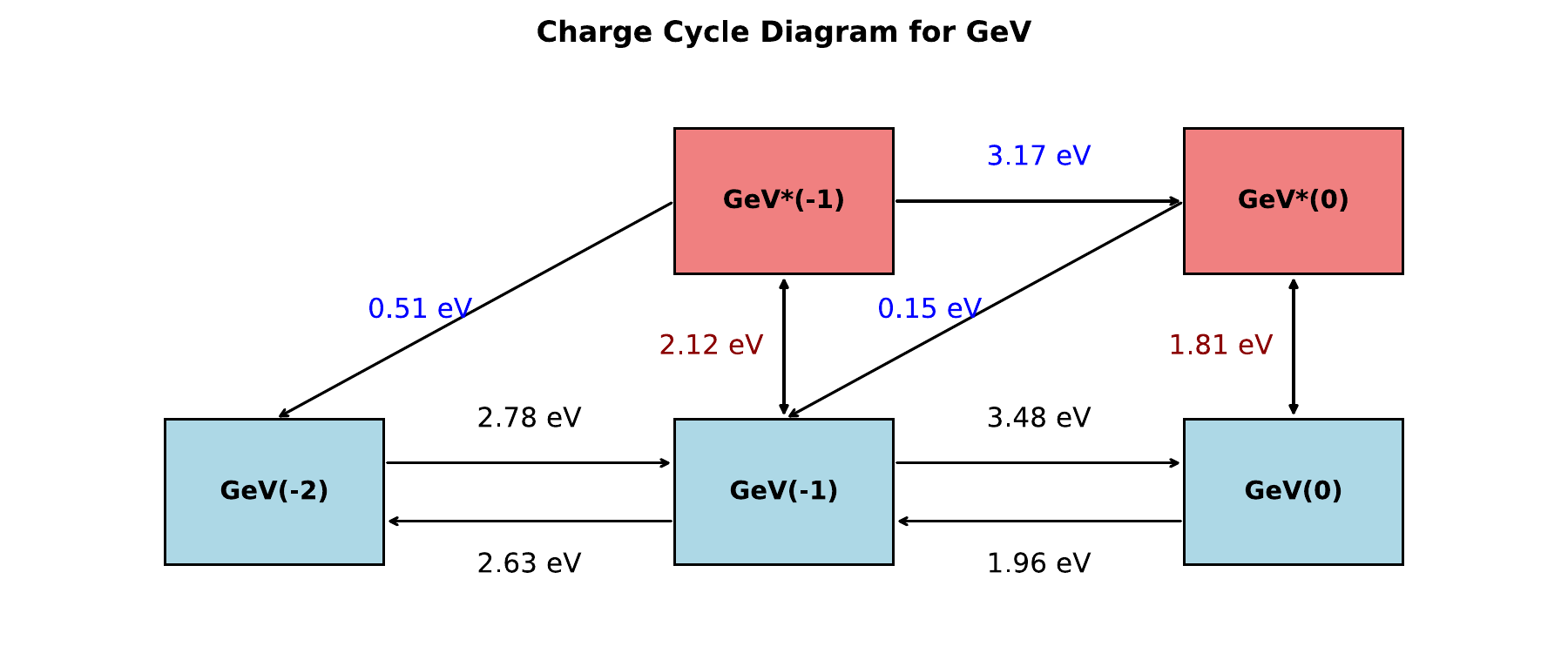}
    \caption{The onset energies for the charge cycle of GeV.}
    \label{fig:GeV}
\end{subfigure}

\vspace{0.5cm}

\begin{subfigure}{0.49\textwidth}
    \centering
    \includegraphics[width=\textwidth]{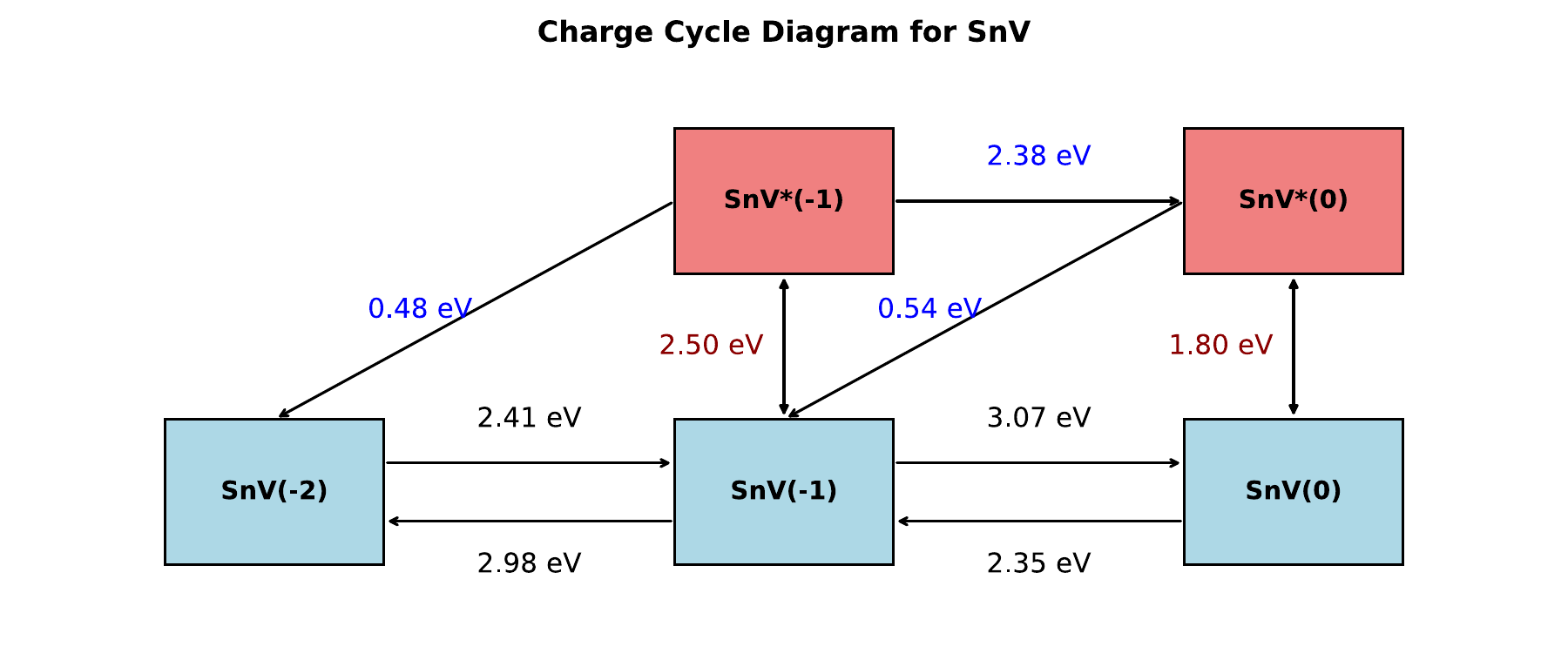}
    \caption{The onset energies for the charge cycle of SnV.}
    \label{fig:SnV}
\end{subfigure}
\hfill
\begin{subfigure}{0.49\textwidth}
    \centering
    \includegraphics[width=\textwidth]{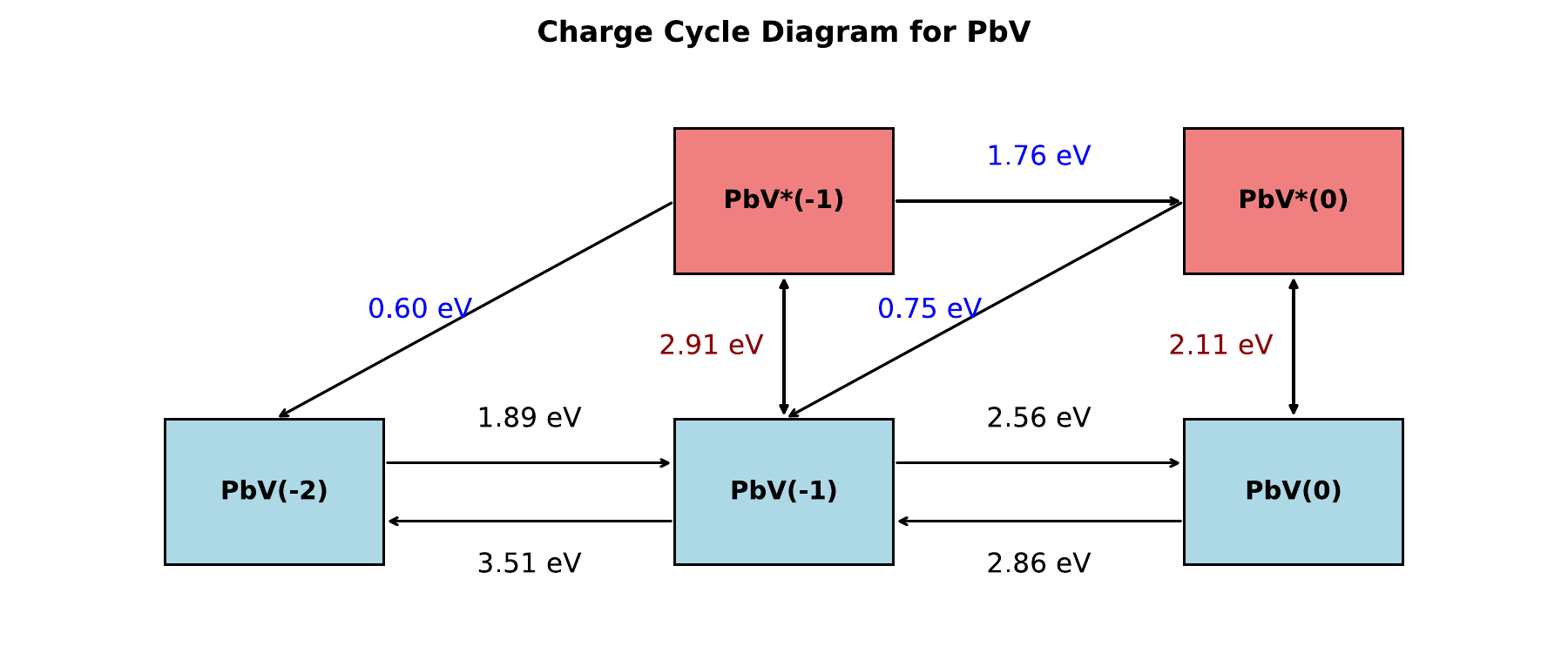}
    \caption{The onset energies for the charge cycle of PbV.}
    \label{fig:PbV}
\end{subfigure}

\caption{Onset energies for the charge cycles of SiV, GeV, SnV, and PbV calculated base on the schemes in Fig. \ref{fig:charge_cycle_scheme_SiV}. The * denotes that defect is in its first excited state.}
\label{fig:charge_cycles}
\end{figure}

In Fig.~\ref{fig:charge_cycles}, the onset energies for each group IV vacancy center are shown, following the schematic in Fig.~\ref{fig:charge_cycle_scheme_SiV}. For the SiV center, we find that photo-ionization to a higher charge state requires photon energies beyond the visible spectrum. This aligns with previous reports in the literature, which state that UV radiation is necessary to restore the neutral charge state~\cite{Charge_state_dyn_Nitrogen, double_charged_SiV}, consistent with our calculations. Furthermore, we determine that when an SiV center is exposed to photons with energies above $1.72$~eV, a transition from $\defectcharge{SiV}{0}$ to $\defectcharge{SiV}{-1}$ becomes possible, enabling subsequent excitation of the excited state of $\defectcharge{SiV}{-1}$. From the excited state, a transition to the $\defectcharge{SiV}{-2}$ state can occur with a low onset energy of $0.45$~eV. While the matrix element for this transition is not yet known, it is reasonable to assume that over time, the concentration of $\defectcharge{SiV}{-2}$ centers will gradually increase. This process is further accelerated when the photon energy exceeds $2.17$~eV, corresponding to the onset energy for ground-state photo-ionization. Unless the photon energy surpasses $3.29$~eV, the other charge states cannot be recovered\footnote{At least, within our model and not through optical means.}

For the other 3 group IV vacancy centers we find that the onset energies are generally lower and therefor it is harder to make prediction about the final charge state. To properly study this we'll need the rate equations. 

% However, it does give us perspective on what is possible in an experiment. For example in \cite{OpticalGatingGeV}, the resonance of the negatively charged GeV center is studied. They find that the $\defectcharge{GeV}{-1}$ reaches a dark state when it is resonantly excited to its first excited state. According to our scheme in Fig.~\ref{fig:GeV} this could possible the $\defectcharge{GeV*}{-1}\rightarrow \defectcharge{GeV}{-2}$ transition, which has small onset energy when compared to the resonant excitation energy. In \cite{OpticalGatingGeV} they use a weak nonresonant laser with a photon energy of $2.33$ eV ($532$ nm) and $3.06$ eV ($405$ nm) to recover the $\defectcharge{GeV}{-1}$. For the laser with photon energy $3.06$ eV the $\defectcharge{GeV}{-2}\rightarrow \defectcharge{GeV}{-1}$ transition is possible according to Fig.~\ref{fig:GeV}. Although, according to our model the photon energy  $2.33$ eV should not be enough for this transition, it is possible that a nearby defect helps the recovery of the  $\defectcharge{GeV}{-1}$ state or that the calculated value is slightly off.

However, it does provide perspective on what is experimentally feasible. For instance, in \cite{OpticalGatingGeV}, the resonance of the negatively charged GeV center is studied. It is found that the $\defectcharge{GeV}{-1}$ state enters a dark state when resonantly excited to its first excited state. According to our scheme in Fig.\ref{fig:GeV}, this could correspond to the $\defectcharge{GeV*}{-1} \rightarrow \defectcharge{GeV}{-2}$ transition, which has a relatively low onset energy compared to the resonant excitation energy. In \cite{OpticalGatingGeV}, a weak nonresonant laser with photon energies of $2.33$ eV ($532$ nm) and $3.06$ eV ($405$ nm) is used to recover the $\defectcharge{GeV}{-1}$ state. For the $3.06$ eV photon energy, the $\defectcharge{GeV}{-2} \rightarrow \defectcharge{GeV}{-1}$ transition is energetically allowed according to Fig.\ref{fig:GeV}. Although our model suggests that $2.33$ eV should not suffice for this transition, it is possible that a nearby defect assists in the recovery of the $\defectcharge{GeV}{-1}$ state \cite{Charge_state_dyn_Nitrogen}, or that the calculated value is slightly inaccurate.

\subsection{Rate equations}
\begin{figure}[H]
\begin{subfigure}{0.49\textwidth}
    \centering
    \includegraphics[width=\textwidth]{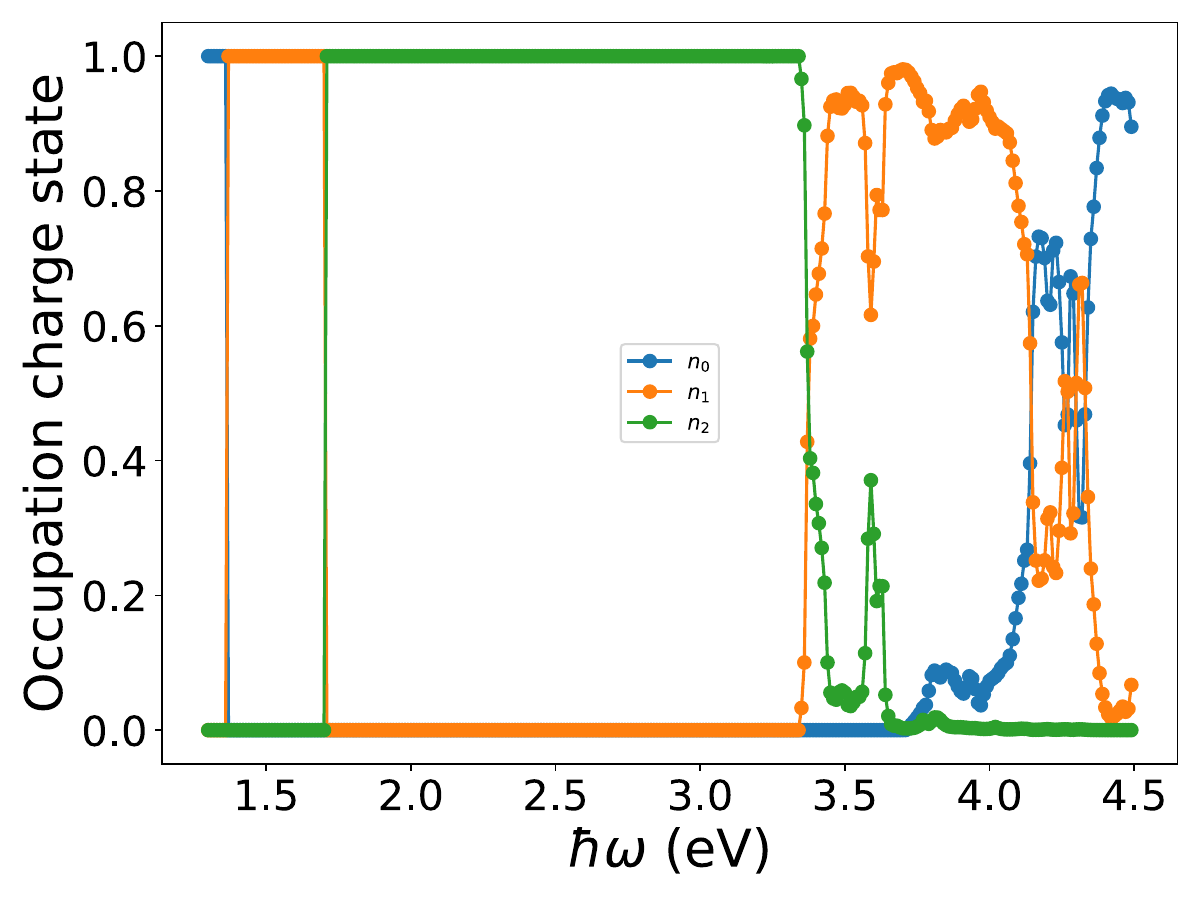}
    \caption{}
    \label{fig:conc_rate_eq_pbe}
\end{subfigure}
\hfill
\begin{subfigure}{0.49\textwidth}
    \centering
    \includegraphics[width=\textwidth]{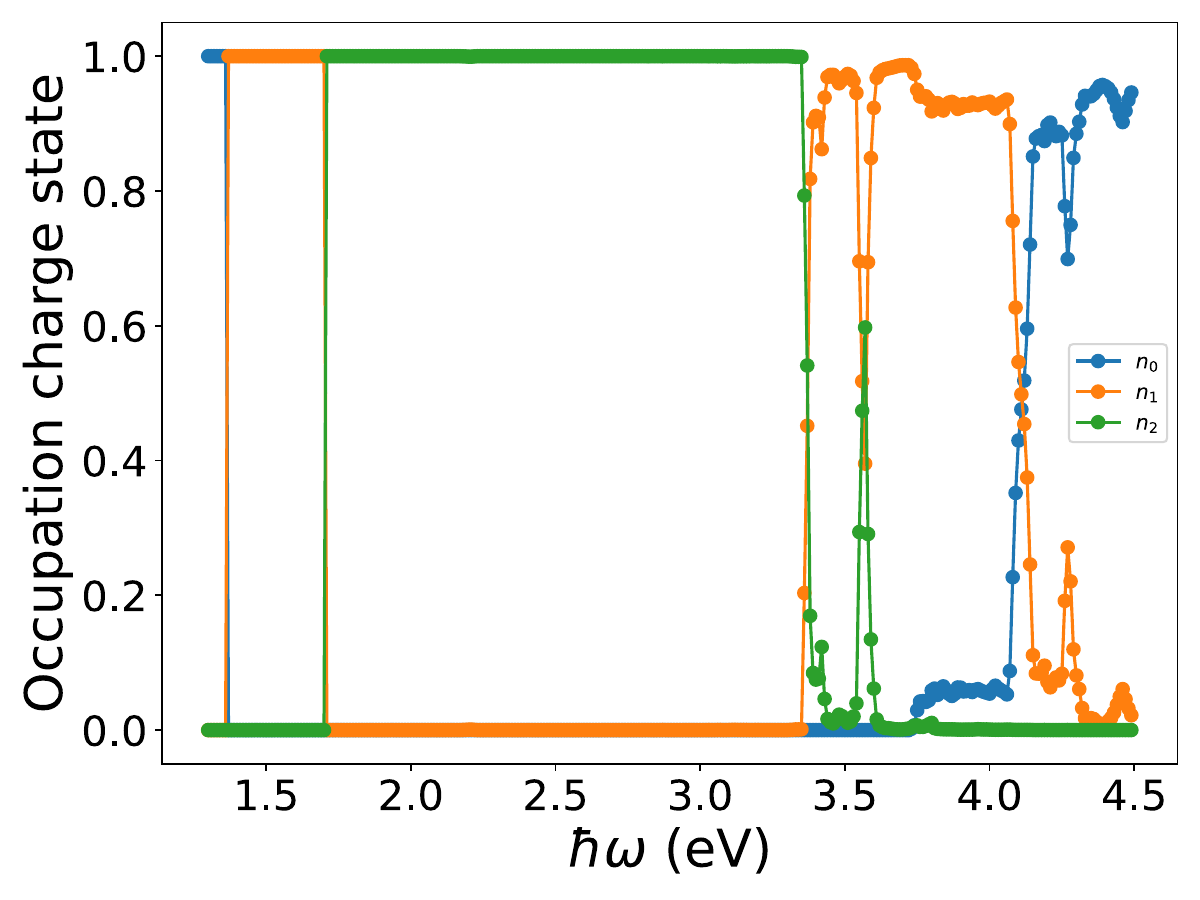}
    \caption{}
    \label{fig:fig:conc_rate_eq_dfthalf}
\end{subfigure}
\caption{The final concentration of the charge states of SiV for different energies of the incident light after a time $1000$ s with a number of photons per unit volume of $N=10^6/\text{cm}^3$. At the beginning of the simulation a the concentration of charge state is assumed to be completely $\defectcharge{SiV}{0}$ as this is the most difficult charge state to achieve. In (a) for the matrix elements determined with PBE and in (b) for the PBE-1/2 matrix elements.}
\label{fig:conc_rate_eq}
\end{figure}

Following the procedure of section \ref{sec:Einstein_coefficients} we determined the rate equation for the silicon vacancy center in diamond. In Fig.~\ref{fig:conc_rate_eq} we see the final concentration of charge states as a function of the incident light. For energies below $3.29$ eV, the onset energy for the transition $\defectcharge{SiV}{-2}\rightarrow \defectcharge{SiV}{-1}$, we find that the final charge state completely the double negative charge state, as the defects could not transition back to a higher charge which was predicted earlier by the onset energy diagram of Fig.~\ref{fig:SiV}. At higher energies we find a mixture of different charge states, where initially $\defectcharge{SiV}{-1}$ dominates after which at around $4.0$ eV the $\defectcharge{SiV}{0}$ concentration start to take over. It should be noted that at such high energies other secondary process which are not included in our model, such as photo-ionization from lower lying defect levels, could start to have a significant role.

\section{Conclusion}

\section*{Acknowledgements}

\printbibliography

% \pagebreak
% \appendix

% \pagebreak

% \appendix

\end{document}